\documentclass{nic-series}
\renewcommand{\mbhead}{\textsf{\tiny Proceedings of the NIC Symposium 2006, Forschungszentrum J\"ulich, 
NIC series vol.\ {\bf 32}, ed.\ by G. M\"unster, D. Wolf, M. Kremer (NIC, J\"ulich, 2006)}}
\begin{document} 
\setcounter{page}{245}
\title{Chain-growth simulations of lattice-peptide adsorption to attractive substrates}
\author{Michael Bachmann \and Wolfhard Janke}
\institute{Institut f\"ur Theoretische Physik, \\
        Universit\"at Leipzig, 04109 Leipzig, Germany\\
        \email{\{bachmann, janke\}@itp.uni-leipzig.de}}
\maketitle
\begin{abstracts}
Based on a newly developed contact-density chain-growth algorithm, we have  
simulated a nongrafted peptide in the vicinity of different attractive substrates.
We analyzed the specificity of the peptide adsorption by focussing on the conformational
transitions the peptide experiences in the binding/unbinding processes. 
In a single simulation run, we obtained the contact density, i.e., the distribution
of intrinsic monomer-monomer contacts and monomer-substrate nearest-neighbor contacts. 
This allows a systematic reweighting to all values of external control 
parameters such as temperature and solvent quality after the simulation. The main result
is the complete solubility-temperature pseudo-phase diagram which is based on the corresponding
specific-heat profile. We find a surprisingly rich structure of pseudo-phases that can roughly be 
classified into compact and expanded conformations in both regimes, adsorption and desorption.
Furthermore, underlying subphases were identified, which, in particular, appear noticeably
in the compact pseudo-phases.     
\end{abstracts}
\section{Introduction}
\label{secintro}
In recent experiments it could be shown that the affinity of peptides to self-assemble
at metal~\cite{brown1} and semiconductor substrates~\cite{whaley1,goede1,willett1} is highly influenced 
by the amino acid content of the peptide, the order of the residues within the sequence, 
the specific substrate, and its crystal orientation at the surface. 

In this study, we investigate the binding specificity with a minimalistic lattice model for the hybrid
system of a peptide in the vicinity of an attractive substrate. Due to the specific properties of the
peptide, this problem is distinguishingly different
from the hybrid system of a (homo)polymer near an adsorbing substrate, which has already been extensively
studied~\cite{vrbova1,singh1,causo1,prellberg1,huang1,bj0}. The peptide sequence consists of hydrophobic
and polar residues, i.e., the 20 protein-building amino acids are classified into only two groups.
The idea behind this hydrophobic-polar (HP) model~\cite{dill1} is that proteins usually possess a compact
hydrophobic core surrounded by a shell of polar residues which screen the core from the aqueous environment.
For this reason and for simplicity, only an effective, short-range attractive force between the hydrophobic monomers
is employed. Furthermore, the peptide is restricted to live on a simple-cubic lattice. The volume exclusion of
the side chains is simply taken into account by considering only self-avoiding linear chains. The energy of
such a lattice peptide is related to the number of hydrophobic nearest-neighbor contacts, $n_{\rm HH}$.

The power of this highly abstract model lies in its simplicity. Peptides with more than 100 residues can 
be studied -- this is 
in striking contrast to refined all-atom protein models, where a systematic analysis of thermodynamic 
properties is only reliably possible for peptides with hardly more than 20 amino acids. It is expected that 
for longer peptides atomic details become less relevant and, therefore, simplified (``coarse-grained'') 
heteropolymer models can give satisfying qualitative answers to specific questions, e.g., regarding
tertiary conformational transitions~\cite{bj1,bj1b}, and systematic sequence analyses~\cite{sbj1}. 
\section{Lattice peptide and hybrid system model}
\label{secmod}
\begin{figure}
\centerline{\epsfxsize=5cm \epsfbox{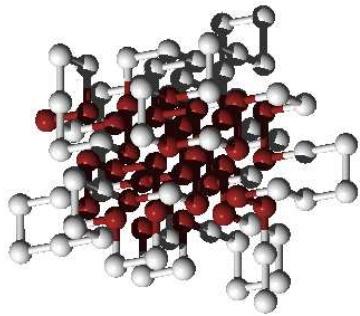}}
\caption{\label{fig:mp} Putative ground-state conformation of the model peptide in the bulk (or in the 
desorbed pseudo-phase).}
\end{figure}

As a model peptide, we use the HP transcription of the 103-residue protein {\em cytochrome c}, which was extensively
studied in the past~\cite{103lat,103toma,hsu1,bj2}. We have first performed a detailed analysis of this  
model peptide in the bulk~\cite{bj2} by applying the newly developed multicanonical chain-growth algorithm~\cite{bj1}. This method
allowed the precise determination of the density of states for this system covering more than 50 orders of magnitude. 
The lowest-energy conformation we identified~\cite{bj2} possesses 56 hydrophobic contacts (see Fig.~\ref{fig:mp}) and 
exhibits a degeneracy of the order of $10^{16}$. It is therefore likely that there still exist lower-lying energetic states. 

Here, this lattice peptide resides in a cavity with
an attractive substrate. In order to study the specificity of residue binding, we distinguish three
substrates with different affinities to attract the peptide monomers: (a) the type-independent attractive, (b) 
the hydrophobic, and (c) the polar substrate. The number of corresponding nearest-neighbor contacts
between monomers and substrate shall be denoted as $n_s^{H+P}$, $n_s^{H}$, and $n_s^{P}$, respectively.
The energy (in arbitrary units) of the hybrid system is then given by 
\begin{equation}
\label{eq:energy}
E_s(n_s,n_{\rm HH})=-n_s-sn_{\rm HH},
\end{equation} 
where $n_s=n_s^{H+P}$, $n_s^{P}$, or $n_s^{H}$, depending on the substrate. 
Besides the temperature $T$, the solubility (or reciprocal solvent 
parameter) $s$ is an external control parameter which governs the quality of the solvent
(the larger the value of $s$, the worse the solvent). The simulation of this model is based on 
a recently developed
contact-density chain-growth algorithm~\cite{bj0} which allows a direct estimation of the degeneracy (or contact density) $g(n_s, n_{\rm HH})$
of macro-states of the system with given contact numbers $n_s$ and $n_{\rm HH}$.  
\section{Contact-density chain-growth algorithm}
\label{secmet}
The contact-density chain-growth algorithm is a suitably enhanced version of the multicanonical chain-growth algorithm~\cite{bj1},
which is based on the pruned-enriched variant~\cite{grassberger2} of the
Rosenbluth chain-growth method~\cite{rosenbluth1}. 
In contrast to move-set based Metropolis Monte Carlo or conventional chain-growth 
methods which would require many separate simulations to obtain 
results for different parameter pairs $(T,s)$ and which frequently suffer from slowing down 
in the low-temperature sector, our method allows the computation of the {\em complete}
contact density for each system within a {\em single} simulation run. Since the contact
density is independent of temperature and solubility, energetic quantities
such as the specific heat can easily be calculated for all values of $T$ and $s$ (nonenergetic
quantities require accumulated densities to be measured within the simulation, but this is also 
no problem). For all systems, 10 independent runs were initialized, each generating $10^8$ conformations. 

In order to regularize the influence of the
unbound conformations and for computational efficiency, the heteropolymer is restricted
to reside in a cage, i.e., in addition to the physically interesting 
attractive surface there is a steric, neutral
wall parallel to it in a distance $z_w$. The value of $z_w$ is chosen sufficiently large 
to keep the influence on the unbound heteropolymer small (in this work we used $z_w=200$).
\section{Pseudo-phase diagram of conformational transitions}
\label{secpseudo}
\begin{figure}
\centerline{\epsfxsize=13cm \epsfbox{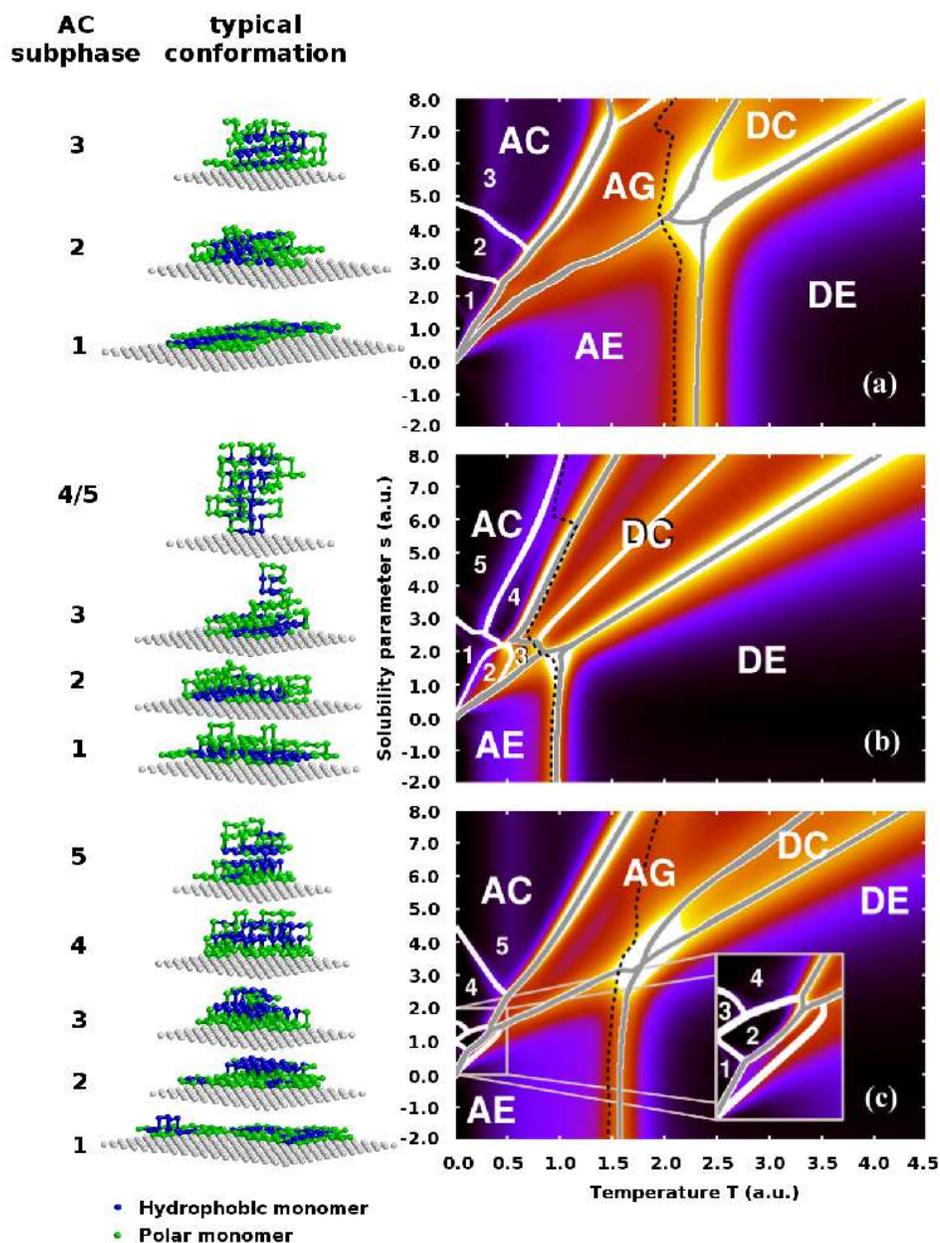}}
\caption{\label{fig:cv} Specific-heat profiles as a function of temperature $T$ and solubility
parameter $s$ of the 103-mer near three different substrates that are attractive for
(a) all, (b) only hydrophobic, and (c) only polar monomers.
White lines indicate the ridges of the profile. Gray lines mark the
main ``phase boundaries''. The dashed black line represents the first-order-like
binding/unbinding transition state, where the contact free energy possesses two minima
(the adsorbed and the desorbed state). In the left panel typical conformations
dominating the associated AC phases of the different systems are shown.}
\end{figure}
Our main interest is devoted to the conformational transitions the peptide experiences in the binding or
adsorption process to the substrates. For a first overview, it is convenient to study the specific heat $C_V$ as
a function of the external parameters temperature $T$ and solubility $s$. Respective ridges and peaks of the
specific heat can be considered as signals of conformational activity. Due to the fixed length of the peptide sequence,
a conventional discussion of thermodynamic phase transitions (e.g., in terms of finite-size scaling) is not possible. 
It should also noted that the behavior of {\em finite} polymer and peptide systems in future nanotechnological 
applications will be of essential interest as a consequence of the need for maximally possible space reduction,
e.g., for nanoelectronic circuits. In such cases, subphase crossover transitions, which are of marginal or
no importance in large systems, strongly influence the self-assembling structure of the polymer or peptide at the 
substrate. 

In Figs.~\ref{fig:cv}(a)-(c) the color-coded profiles of the specific heats for the different
substrates are shown (the brighter the colour,
the larger the value of $C_V$). We interpret the ridges (for accentuation marked
by white and gray lines) as the boundaries of the pseudo-phases. It should be noted, however, 
that in such a finite system the exact positions of active regions exhibited by fluctuations of other quantities 
usually deviate, but the qualitative behavior is similar.~\cite{bj1} 
Despite the surprisingly rich and complex phase behavior there are main ``phases'' that can be
distinguished in all three systems. 
These are separated in Figs.~\ref{fig:cv}(a)-(c) by gray lines.
Comparing the three systems we find that they all possess
pseudo-phases, where adsorbed compact (AC), adsorbed expanded (AE), desorbed compact (DC), and
desorbed expanded (DE) conformations dominate, similar to the generic phase diagram of a
homopolymer~\cite{bj0}. ``Compact'' here 
means that the heteropolymer has formed a 
dense hydrophobic core, while expanded conformations form dissolved, random-coil-like structures.
The sequence and substrate specificity of
heteropolymers generates, of course, new interesting and selective phenomena
not available for homopolymers. One example is the pseudo-phase of adsorbed globules (AG), which is noticeably present
only in those systems, where all monomers are equally attractive to the substrate (Fig.~\ref{fig:cv}(a)) and
where polar monomers favour contact with the surface (Fig.~\ref{fig:cv}(b)). In this phase, the conformations
are intermediates in the binding/unbinding region. This means that monomers 
currently desorbed from the substrate have not yet found their position within a compact conformation.

The strongest difference between the three systems is their behavior in pseudo-phase AC, which is
roughly parameterized by $s>5T$. Representative conformations for all AC subphases are 
shown in the left panel of Fig.~\ref{fig:cv}. 
If hydrophobic and polar monomers are equally attracted by the substrate (Fig.~\ref{fig:cv}(a)),
we find three AC subphases in the parameter space plotted. In AC1 film-like conformations
dominate, i.e., all 103 monomers are in contact with the substrate. 
The formation of a single, compact hydrophobic core proceeds by layering transitions from AC1 to
AC3 via AC2. The reason for the existence of phase AC2 is the reduced cooperativity of
the polar monomers due to their surface attraction. In AC3,
the heteropolymer has maximized the number of hydrophobic contacts and only local arrangements of monomers 
on the surface of the very compact structure lead to the still possible maximum number of substrate
contacts. 

The AC heteropolymer conformations adsorbed at a surface that is only attractive to
hydrophobic monomers (Fig.~\ref{fig:cv}(b)) depend on two concurring 
hydrophobic forces: substrate attraction and formation of intrinsic
contacts. 
The {\em single} film-like hydrophobic domain in AC1 is 
maximally compact, at the expense of displacing polar monomers into upper layers.
In subphase AC2 intrinsic hydrophobic contacts are entropically 
broken, while 
AC3 exhibits hydrophobic layers at the expense of hydrophobic substrate contacts. A dramatic, highly 
cooperative, hydrophobic collapse accompanies the transitions from AC1 to AC4/5, where in a one-step process 
the compact two-dimensional domain transforms to the compact
three-dimensional hydrophobic core. 

Not less exciting is the subphase structure of the heteropolymer interacting with a polar substrate
(Fig.~\ref{fig:cv}(c)).
For small values of $s$ and $T$, the behavior of the heteropolymer is dominated by the concurrence
between polar monomers contacting the substrate and hydrophobic monomers favouring the formation
of a hydrophobic core, which, however, also requires cooperativity of the polar monomers.
In AC1, film-like conformations 
with disconnected hydrophobic clusters dominate. Entering AC2, a second hydrophobic layer forms at the expense
of a reduction of polar substrate contacts. 
In contrast to the case of a hydrophobic substrate (Fig.~\ref{fig:cv}(b)), the strong surface attraction of 
polar monomers hinders here the formation of a compact hydrophobic core (AC2/3 to AC5) which results
in the intermediate subphase AC4.
\begin{figure}
\centerline{\epsfxsize=13cm \epsfbox{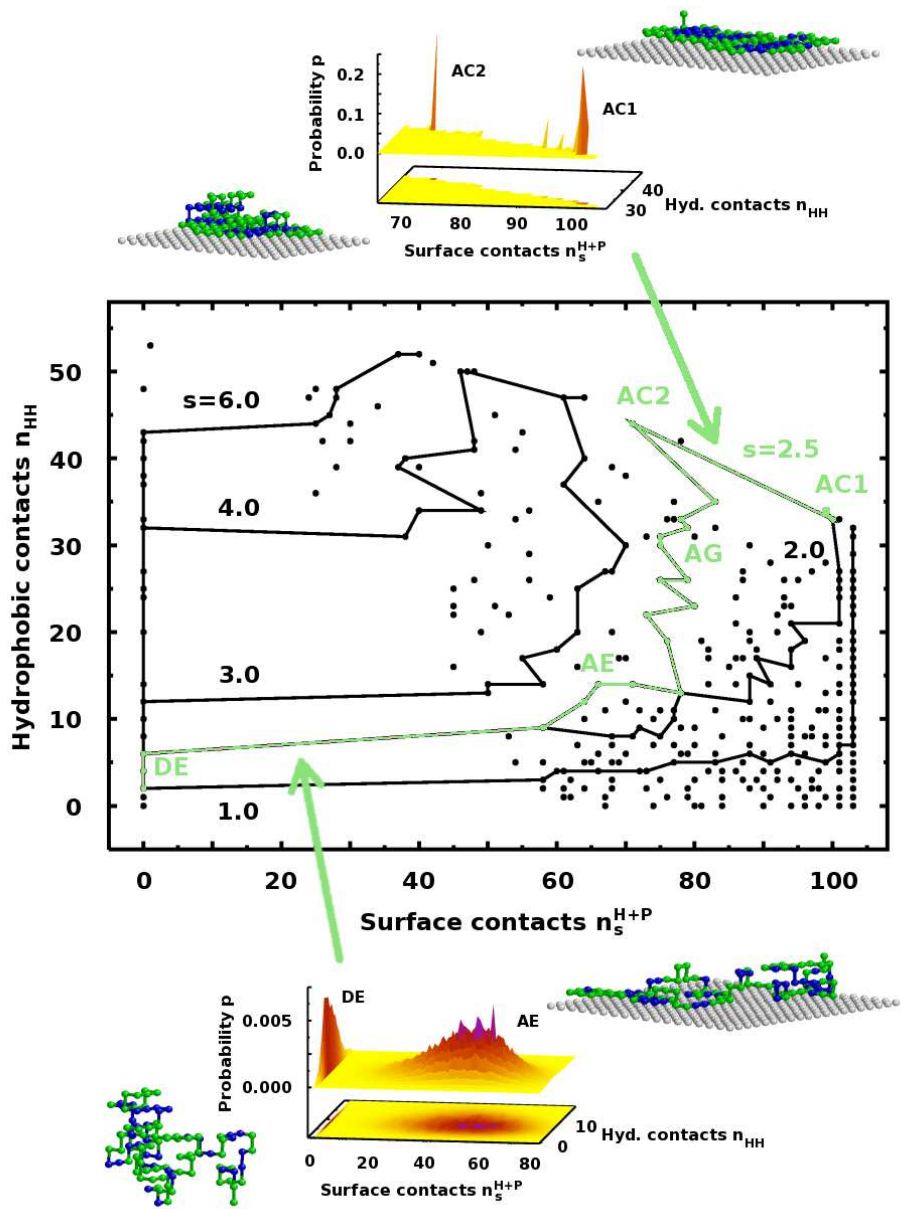}}
\caption{\label{fig:lm} Contact-number map of all free-energy minima for the 103-mer
and substrate equally attractive to all monomers. Full circles correspond to minima of the contact free energy 
$F_{T,s}(n_s^{\rm H+P},n_{\rm HH})$
in the parameter space $T\in [0,10]$, $s\in[-2,10]$. Lines illustrate how the contact free energy changes
with the temperature at constant solvent parameter $s$. For the exemplified
solvent with $s=2.5$, the peptide experiences near $T=0.35$ a sharp first-order-like layering transition
between single- to double-layer conformations (AC1,2). Passing the regimes of adsorbed globules (AG)
and expanded conformations (AE), the discontinuous binding/unbinding transition from AE to DE happens
near $T=2.14$. In the DE phase the ensemble is dominated by desorbed, expanded conformations.
Representative conformations of the phases are shown next to the respective peaks of the probability
distributions.
}
\end{figure}
\section{Free-energy landscape from a different perspective}
\label{secfree}
The contact numbers $n_s$ and $n_{\rm HH}$ are kind of order parameters adequately describing the 
macro-state of the system. With its degeneracy  $g(n_s,n_{\rm HH})$, we define the contact free energy as 
$F_{T,s}(n_s,n_{\rm HH}) \sim -T\ln\, g(n_s,n_{\rm HH})\exp(-E_s/T)$ and 
the probability for a macro-state with $n_s$ substrate and $n_{\rm HH}$ hydrophobic contacts as
$p_{T,s}(n_s,n_{\rm HH})\sim g(n_s,n_{\rm HH})\exp(-E_s/T)$. 
Assuming that the minimum of the free-energy landscape $F_{T,s}(n_s^{(0)},n_{\rm HH}^{(0)})\to {\rm min}$ for 
given external parameters $s$ and $T$ is related to the class of macro-states with $n_s^{(0)}$ surface and 
$n_{\rm HH}^{(0)}$ hydrophobic contacts, this class dominates the phase the system resides in. For this
reason, it is instructive to calculate all minima of the contact free energy and to determine the 
associated contact numbers in a wide range of values for the external parameters. The map of all possible free-energy 
minima in the range of external parameters $T\in[0,10]$ and $s\in[-2,10]$ is shown in Fig.~\ref{fig:lm} for
the peptide in the vicinity of a substrate that is equally attractive for both hydrophobic and polar monomers. 
Solid lines visualize ``paths'' through the free energy landscape when changing temperature under constant solvent 
($s={\rm const}$) conditions. Let us follow the exemplified trajectory for $s=2.5$. 
Starting at very low temperatures, 
we know from the pseudo-phase diagram in Fig.~\ref{fig:cv}(a) that the system resides in pseudo-phase AC1. This means
that the macro-state of the peptide is dominated by the class of compact, film-like single-layer conformations. The system
obviously prefers surface contacts at the expense of hydrophobic contacts. Nonetheless, the formation of compact
hydrophobic domains in the two-dimensional topology is energetically favored but maximal compactness is hindered by the steric
influence of the substrate-binding polar residues. Increasing the temperature, the system experiences close to $T\approx 0.35$
a sharp first-order-like conformational transition, and a second layer forms (AC2). This is a mainly entropy-driven transition
as the extension into the third dimension perpendicular to the substrate surface increases the number of possible 
peptide conformations. Furthermore, the loss of energetically favored substrate contacts of polar monomers is 
partly compensated by the energetic gain due to the more compact hydrophobic domains. Increasing the temperature
further, the density of the hydrophobic domains reduces and overall compact conformations dominate in the globular
pseudo-phase AG. Reaching AE, the number of hydrophobic contacts decreases further, and also the total number of
substrate contacts. Extended, dissolved conformations dominate. The transitions from AC2 to AE via AG are comparatively 
``smooth'', i.e., no immediate
changes in the contact numbers passing the transition lines are noticed. Therefore, these conformational transitions
could be classified as second-order-like. The situation is different when approaching the unbinding
transition line from AE  close to $T\approx 2.14$. This transition is accompanied by a dramatic loss of substrate 
contacts -- the peptide desorbs from the substrate and behaves in pseudo-phase DE like a free peptide, i.e., only the substrate
and the opposite neutral wall regularize the translational degree of freedom perpendicular to the walls, but rotational
symmetries are unbroken (at least for conformations not touching one of the walls). As the probability distribution
in Fig.~\ref{fig:lm} shows, the unbinding transition is also first-order-like, i.e., close to the transition line, there
is a coexistence of adsorbing and desorbing classes of conformations.  

\section{Concluding Remarks}
\label{secconc}
Summarizing, we have performed a detailed analysis of the pseudo-phase diagrams in the $T$-$s$ plane for
a selected heteropolymer with 103 monomers in 
cavities with an adsorbing substrate being either attractive independently of the monomer type,
or selective to hydrophobic or polar monomers, respectively. 
Although our model is very simple and the focus is on hydrophobic and polar effects only,  
we find, beyond the expected
adsorbed and desorbed phases, a rich subphase structure in
the adsorbed phases. In these regions, the substrate-specificity depends in detail on the 
quality of the solvent.

Since current experimental equipment is capable to reveal molecular structures at 
the nanometer scale, it should be possible to investigate the grafted structures
dependent on the solvent quality. 
This is essential for answering the question under what circumstances 
binding forces are strong enough to refold peptides or proteins. The vision of future 
biotechnological and medical applications is fascinating as it ranges from protein-specific 
sensory devices to molecular electronic devices at the nanoscale.
\section*{Acknowledgments}
This work is partially supported by a DFG (German Science Foundation) grant
under contract No.\ JA 483/24-1. We thank the John von Neumann Institute for 
Computing (NIC), Forschungszentrum
J\"ulich, for providing access to their 
supercomputer JUMP under grant No.\ hlz11.

\end{document}